\documentclass[12pt]{article}
\usepackage{amssymb}
\usepackage{amsmath, epsfig}

\usepackage{color}

\newcommand{\be}{\begin{equation}}
\newcommand{\en}{\end{equation}}

\renewcommand{\vec}[1]{\boldsymbol{#1}}
\newcommand{\ii}{\textrm{i}}
\newcommand{\ee}{\textrm{e}}
\newcommand{\demi}{\textstyle{\frac{1}{2}}}

\begin{document}

\title{Third- and fourth-order elasticity of biological soft tissues}


\author{Michel Destrade$^a$, Michael D. Gilchrist$^a$, Raymond W. Ogden$^b$ \\[12pt]
$^a$School of Electrical, Electronic, and Mechanical Engineering,\\
University College Dublin,\\ Belfield, Dublin 4, Ireland.\\[12pt]
$^b$Department of Mathematics,\\
University of Glasgow,\\
University Gardens,
Glasgow G12 8QW,
Scotland, UK.}

\date{}
\maketitle

\begin{abstract}

In the theory of weakly non-linear elasticity, Hamilton et al. [J. Acoust. Soc. Am. \textbf{116} (2004) 41] identified $W = \mu I_2 + (A/3)I_3 + D I_2^2$ as the fourth-order expansion of the strain-energy density for incompressible isotropic solids. Subsequently, much effort focused on theoretical and experimental developments linked to this expression in order to inform the modeling of gels and soft biological tissues. However, while many soft tissues can be treated as incompressible, they are not in general isotropic, and their anisotropy is associated with the presence of oriented collagen fiber bundles. Here the expansion of $W$ is carried up to fourth-order in the case where there exists one family of parallel fibers in the tissue. The results are then applied to acoustoelasticity, with a view to determining the second- and third-order nonlinear constants by employing small-amplitude transverse waves propagating in a deformed soft tissue.

\end{abstract}
\newpage




\section{INTRODUCTION}


Due to their high fluid content, many soft biological tissues and gels exhibit nearly \emph{incompressible} behavior: they are constrained to undergo essentially volume-preserving deformations and motions.  This has long been known and, in fact, can traced back to the experiments of Jan Swammerdam\cite{Swam58, Cobb02} in 1758. Mathematically,  the internal constraint of incompressibility is expressed by imposing the condition $\det \vec{F} = 1$ at all times, where $\vec{F}$ is the deformation gradient.
In \emph{exact} nonlinear isotropic elasticity theory, the strain-energy density $W$ is written in fullest generality as a function of $I$, $II$, $III$, the first three principal invariants of the right Cauchy-Green strain tensor $\vec{C} = \vec{F}^T\vec{F}$, which are defined by
\be \label{III}
I = \text{tr} \ \vec{C}, \qquad
II = \demi [(\text{tr}\ \vec{C})^2 - \text{tr}(\vec{C}^2)], \qquad
III = \det \vec{C} = (\det\vec{F})^2.
\en
It is then easy to enforce incompressibility, which requires that $III=1$ at all times and hence that $W = W(I, II)$ only.

In \emph{weakly} nonlinear isotropic elasticity theory, $W$ is expanded in terms of invariants of the Green-Lagrange strain tensor $\vec{E}  = \demi (\vec{C} - \vec{I})$, either using the set
\be \label{i_k}
i_1 = \text{tr} \ \vec{E}, \qquad
i_2 = \demi [(\text{tr} \ \vec{E})^2 - \text{tr}(\vec{E}^2)], \qquad
i_3 = \det \vec{E},
\en
(this is the choice of Murnaghan\cite{Murn51})
or the set
\be \label{I_k}
I_1 = \text{tr} \ \vec{E}, \qquad
I_2 = \text{tr}\ \vec{E}^2, \qquad
I_3 = \text{tr} \ \vec{E}^3,
\en
(this is the choice of Landau and Lifshitz\cite{LaLi86}).  In each case the respective terms are of orders 1, 2 and 3 in $\vec{E}$.

Then, in \emph{third-order} elasticity theory, $W$ is expanded as a linear combination of $i_1$, $i_2$, $i_3$, $i_1 i_2$, and $i_1^3$ (or, equivalently,  $I_1$, $I_2$, $I_3$, $I_1 I_2$, and $I_1^3$) and higher-order terms are neglected --- and thus, 5 elastic constants are required to describe third-order solids. [Note that some authors refer to this as second-order elasticity since the stress is second order in the strain; a similar comment applies to fourth-order versus third-order elasticity.]
In \emph{fourth-order} elasticity theory, terms linear in $i_1^4$, $i_2^2$, $i_1 i_3$, and $i_1^2 i_2$ (or, equivalently,  $I_1^4$, $I_2^2$, $I_1 I_3$, and $I_1^2 I_2$) are also retained --- and thus, 9 elastic constants are required to describe fourth-order solids.
Now, however, it is slightly more complicated to enforce incompressibility, because the constraint involves a combination of invariants of $\vec{E}$, rather than a single invariant of $\vec{C}$.
Nevertheless, this can be achieved because of obvious connections between the sets \eqref{III}, \eqref{i_k}, and \eqref{I_k}. This was done long ago by Rivlin and Saunders\cite{RiSa51} in 1951, and by Ogden\cite{Ogde74} in 1974, for third- and fourth-order incompressible solids, respectively. In particular, these authors showed  that the number of elastic constants drops from 5 to 2 for third-order solids, and from 9 to 3 for fourth-order solids compared with the compressible case.

Recently, Hamilton et al.\cite{HaIZ04}, unaware of these previous contributions, rediscovered the latter result, and wrote $W$ in the form
\be \label{incomp0}
W = \mu I_2  + \frac{A}{3}I_3  + D I_2^2,
\en
where $\mu$, $A$, and $D$ are the respective second, third-, and fourth-order elasticity constants of weakly nonlinear incompressible isotropic elasticity.
Subsequently, several papers\cite{HaIZ04B, HaIZ07, Genn07, Reni07, JCGB07, OSIR07, VaPo07, Woch08, Reni08b, Reni08a, WoHI08, Ostro08, MKPC08, Doma09, MiPK09, ZaIH09} have been devoted to the study of the theoretical and experimental implications of this expansion for linearized and nonlinear waves and vibrations in incompressible solids.
Experiments have been conducted on isotropic gels and phantoms to measure the nonlinear elastic coefficients, and good progress has been achieved in the understanding and implementation of elastographic techniques\cite{Genn07, Reni07, JCGB07, Reni08b, Reni08a}.

Now, the ultimate goal of elastography is the imaging of biological soft tissues, and it is therefore important to be able to evaluate $\mu$, $A$, and $D$ experimentally.
At the same time, it must be kept in mind that soft tissues are \emph{an}isotropic solids, in contrast to gels and phantoms.
This anisotropy is due to the presence of oriented collagen fiber bundles, which are three orders of magnitude stiffer than the surrounding tissue in which they are embedded (this surrounding tissue consists of a network of elastin fibers, smooth muscle cells and proteoglycans, inter alia, which collectively form an essentially isotropic `matrix'.)
The natural conclusion of this consideration is that there is a need for expansions of $W$ up to the third or, depending on the context, fourth order for incompressible anisotropic solids.
For the particular application considered in Section III of the present paper, however, it suffices to pursue the analysis only up to the third order.


\section{FOURTH-ORDER TRANSVERSELY \\ ISOTROPIC SOLIDS}


Here we consider soft tissues with one preferred direction, associated with a family of parallel fibers of collagen.
We denote by $\vec{A}$ the unit vector in that direction when the solid is unloaded and at rest.
The theory of Spencer\cite{Spen72} tells us that in all generality, $W$ is a function of at most 5 invariants in exact nonlinear compressible elasticity theory.
One valid choice for these quantities is
\be \label{comp}
W = W(I_1, I_2, I_3, I_4, I_5), \qquad \text{where} \qquad I_4 \equiv \vec{A}\cdot \vec{E A}, \quad I_5 \equiv \vec{A} \cdot \vec{E}^2 \vec{A}.
\en
Now we expand $W$ up to terms of order 4 in the Green-Lagrange strain.
This involves a linear combination of the following quantities:
\begin{align} \label{compr}
& \text{order 2: } I_1^2, \ I_2, \ I_4^2, \ I_5, \ I_1 I_4, \\ \notag
& \text{order 3: } I_1^3, \ I_1 I_2, \ I_3, \ I_1 I_4^2, \ I_1 I_5, \ I_1^2 I_4, \ I_2 I_4, \ I_4^3, \ I_4 I_5, \\ \notag
& \text{order 4: } I_1^4, \ I_1^2 I_2, \ I_1 I_3, \ I_2^2, \ I_1 I_4^3, \ I_1^2 I_5, \ I_1^2 \ I_4^2, \ I_1^3 I_4, \\ \notag
& \phantom{\text{order 4: } } \ I_2 I_4^2, \ I_2 I_5, \ I_4^4, \ I_5^2, \ I_1 I_2 I_4, \ I_1 I_4 I_5, \ I_3 I_4.
\end{align}
Note that the first-order terms are omitted from the list because they give rise to constant stresses.
Hence there are 29 elastic constants for fourth-order, transversely isotropic, compressible solids.

In incompressible solids, however, the isotropic invariants are not independent.
They are connected by the incompressibility condition, which can be written in terms of the invariants \eqref{i_k} exactly as\cite{Ogde74}
$i_1=-2i_2-4i_3$, or equivalently, in terms of the invariants \eqref{I_k}, as
\be
I_1 = I_2  - \tfrac{4}{3} I_3-I_1^2+2I_1I_2-\tfrac{2}{3}I_1^3,
\en
so that the number of independent quantities in \eqref{compr} is greatly reduced.
For instance, $I_1^2$ is of fourth order, $I_1 I_2 = I_2^2$ and $I_1^3=0$ at fourth-order.
This allows the list to be shortened to
\begin{align} \label{incompr}
& \text{order 2: } I_2, \ I_4^2, \ I_5, \\ \notag
& \text{order 3: } I_3, \ I_2 I_4, \ I_4^3, \ I_4 I_5, \\ \notag
& \text{order 4: } I_2^2, \ I_2 I_4^2, \ I_2 I_5, \ I_4^4, \ I_5^2, \ I_3 I_4,
\end{align}
which are associated with 13 elastic constants \emph{in toto}.

For third-order, transversely isotropic, incompressible, nonlinear elastic solids, for example, the strain-energy density is thus expressible in the form
\be \label{3rdTI}
W = \mu I_2  + \frac{A}{3}I_3  + \alpha_1 I_4^2 + \alpha_2 I_5 + \alpha_3 I_2 I_4 + \alpha_4 I_4^3 + \alpha_5 I_4 I_5,
\en
where $\mu$, $\alpha_1$,  $\alpha_2$ are second-order elastic constants and $A$, $\alpha_3$, $\alpha_4$, $\alpha_5$,  $\alpha_6$ are third-order elastic constants. A lengthier expansion is required at fourth order, which requires a linear combination of the six additional terms listed in \eqref{incompr}. We do not include these here. Indeed, it suffices for our present purposes to restrict attention to second- and third-order terms in $W$ since at this order the speed of infinitesimal plane waves depends linearly on the strain through the third-order constants, even for the specialization to an isotropic material.


\section{ACOUSTOELASTICITY}


We now consider the propagation of small-amplitude plane body waves in a deformed soft tissue with one family of parallel fibers. We assume that the solid has been pre-stressed by the application of the Cauchy stress $\vec{\sigma}$, say, giving rise to a pre-deformation with corresponding  deformation gradient $\vec{F}$.

Let $\vec{u} = \vec{u}(\vec{x},t)$ denote the mechanical displacement associated with the wave motion, $\rho$ the (constant) mass density, and $p$ the incremental Lagrange multiplier due to the incompressibility constraint.
Then the incremental equations of motion and of incompressibility read\cite{Ogde84}
\be \label{mtn}
\mathcal{A}_{0piqj} u_{j,pq} - p_{,i} = \rho \partial^2 u_i/\partial t^2, \qquad u_{i,i} =0,
\en
respectively,
where the comma signifies partial differentiation with respect to the current coordinates $\vec{x}\equiv x_i$ and $\vec{\mathcal{A}}_{\vec{0}}$ is the fourth-order tensor of instantaneous moduli, with components
\be
\mathcal{A}_{0piqj} = F_{p\alpha} F_{q\beta}\delta_{i j} \dfrac{\partial W}{\partial E_{\alpha \beta}} + F_{p\alpha} F_{q\beta} F_{j\nu} F_{i\gamma}\dfrac{\partial^2 W}{\partial E_{\alpha \gamma} \partial E_{\beta \nu}}.
\en
This expression is exact,  while to compute the derivatives of $W$ in respect of \eqref{3rdTI} the following quantities are needed:
\begin{align}
& \dfrac{\partial I_2}{\partial E_{\alpha\beta}} = 2 E_{\alpha \beta},
\notag \\
& \dfrac{\partial^2 I_2}{\partial E_{\alpha \gamma} \partial E_{\beta \nu}} = \delta_{\alpha \beta}\delta_{\gamma\nu} + \delta_{\alpha\nu} \delta_{\gamma\beta},
\notag \\
& \dfrac{\partial I_3}{\partial E_{\alpha\beta}} = 3 E_{\alpha \gamma} E_{\gamma \beta}, 
\notag \\
&  \dfrac{\partial^2 I_3}{\partial E_{\alpha \gamma} \partial E_{\beta \nu}} = \frac{3}{2}\left( \delta_{\alpha \beta} E_{\nu \gamma} + \delta_{\alpha\nu}E_{\beta\gamma} + \delta_{\gamma\nu} E_{\alpha\beta} + \delta_{\gamma\beta}E_{\alpha\nu}\right),
\notag \\
& \dfrac{\partial I_4}{\partial E_{\alpha\beta}} = A_{\alpha} A_{\beta},
\notag \\
&  \dfrac{\partial^2 I_4}{\partial E_{\alpha \gamma} \partial E_{\beta \nu}} = 0,
\notag \\
& \dfrac{\partial I_5}{\partial E_{\alpha\beta}} = A_{\alpha}A_{\gamma} E_{\beta\gamma} + A_\beta A_\gamma E_{\alpha \gamma},
\notag \\
&  \dfrac{\partial^2 I_5}{\partial E_{\alpha \gamma} \partial E_{\beta \nu}} = \frac{1}{2} (A_\alpha A_\nu \delta_{\beta \gamma} + A_\alpha A_\beta \delta_{\gamma\nu} + A_\beta A_\gamma \delta_{\alpha\nu} + A_\gamma A_\nu \delta_{\alpha\beta}).
\end{align}

Let us now consider the propagation of homogeneous plane body waves, in the form
\be
\vec{u} = \vec{a} \ee^{\ii k (\vec{n \cdot x} - v t)}, \qquad
p = \ii k P \ee^{\ii k (\vec{n \cdot x} - v t)},
\en
where $\vec{a}$ is the unit vector in the direction of linear polarization, $k$ is the wave number, $v$ is the phase speed, and $P$ is a scalar.
Then the incremental incompressibility condition \eqref{mtn}$_2$ gives: $\vec{a}\cdot\vec{n}=0$, and the wave is thus \emph{transverse}.
The incremental equations of motion \eqref{mtn}$_1$ can be written in the form
\be
\vec{Q}(\vec{n})\vec{a} - P\vec{n} = \rho v^2 \vec{a},
\qquad
\text{where} \qquad
[\vec{Q}(\vec{n})]_{ij} = \mathcal{A}_{0piqj}n_p n_q.\label{Q-P}
\en
Taking the dot product of this equation with $\vec{n}$ gives an expression for $P$.
Substituting this back into \eqref{Q-P} and using the orthogonality of the propagation and polarization vectors,
we end up with the symmetric eigenvalue problem,
\be
[\vec{I} - \vec{n}\otimes\vec{n} ] \vec{Q}(\vec{n})[\vec{I} - \vec{n}\otimes\vec{n} ]\vec{a} = \rho v^2 \vec{a}
\en
for the wave speed and polarization for any given direction of propagation.
The wave speed $v$ is given simply by $\rho v^2 = \mathcal{A}_{0piqj}n_p n_q a_i a_j$.

As a simple example of the application of the above equations, consider the case where a sample of soft tissue is under uniaxial tension or compression with the direction of tension parallel to the fibers.
We denote by $e$ the elongation of the sample in the direction ($x_1$, say) of the uniaxial stress and discard all terms equal to or higher than $e^2$ in the expansions.
In particular, we obtain the approximations
\be
\vec{F} = \text{diag}[1+e, 1-e/2, 1-e/2], \qquad
\vec{E} =  \text{diag}[e, -e/2, -e/2].
\en
Also, we find that there are only 15 non-zero components of $\vec{\mathcal{A}_0}$ in the ($x_1, x_2, x_3$) coordinate system because, in this special case, the principal axes of stress and strain are aligned.
Now let $\theta$ be the angle between the direction of propagation and the $x_1$ axis.
Then the \emph{secular equation}, which gives the wave speed in terms of the elastic moduli, has the same form as in the isotropic case\cite{Ogde07},
with the difference that the moduli depend not only on the isotropic elastic constants $\mu$ and $A$, but also on the anisotropic elastic constants $\alpha_1$, $\alpha_2$, $\alpha_3$, $\alpha_4$, and $\alpha_5$.
It reads\cite{Ogde07}
\be \label{bulk}
\rho v^2 = (\alpha + \gamma - 2\beta)\cos^4\theta + 2(\beta - \gamma)\cos^2\theta + \gamma,
\en
where
\be
\alpha = \mathcal{A}_{01212}, \qquad
2\beta = \mathcal{A}_{01111} + \mathcal{A}_{02222} - 2\mathcal{A}_{01122} - 2 \mathcal{A}_{01221},
\qquad
\gamma = \mathcal{A}_{02121}.
\en
Lengthy, but straightforward, calculations reveal that
\begin{align}
& 2 \beta - \alpha -\gamma = 2\alpha_1 + 2(4\alpha_1 + 3 \alpha_2 + 3\alpha_3 + 3 \alpha_4 + 2\alpha_5)e,
\\ \notag
& 2(\beta - \gamma)  = 2 \alpha_1 + (3\mu + 10\alpha_1 + 8 \alpha_2 +  6\alpha_3 + 6 \alpha_4 + 4\alpha_5)e,
\\ \notag
& 4\gamma = 4\mu + 2\alpha_2 + (A + 2\alpha_2 + 4\alpha_3 + 2\alpha_5)e.
\end{align}
Clearly, Eq. \eqref{bulk} provides a direct means of evaluating the elastic constants, by measuring the transverse wave speed for a variety of $e$ and $\theta$ combinations.


\section{CONCLUSION}


We have shown that 7 elastic constants are required to describe third-order incompressible solids with one family of parallel fibers, and we have indicated that these constants may be determined by using the acoustoelastic effect.
We have also noted that 13 constants are needed for fourth-order solids.

In closing this Letter, we remark that although many biological soft tissues exhibit transverse isotropy (one family of parallel fibers), many more must be modeled as orthotropic materials (two families of parallel fibers).
Then, the list of invariants is increased from 4 ($I_1$, $I_2$, $I_4$, $I_5$) to 7, with the following 3 additions:
\be
I_6 \equiv \vec{B}\cdot \vec{EB}, \qquad
I_7 \equiv \vec{B}\cdot \vec{E}^2 \vec{B}, \qquad
I_8 \equiv (\vec{A}\cdot \vec{EB})\vec{A}\cdot\vec{B},
\en
where $\vec{B}$ is the unit vector in the direction of the second family of parallel fibers.
In that case, an expansion of $W$ up to third- or fourth-order yields such a large number of elastic constants to be determined, that it defeats its own usefulness.
We would argue that it is more advantageous to turn to the exact nonlinear theory, where constitutive anisotropic models have been successfully evaluated against experimental test data to capture well the mechanics of orthotropic soft tissues.
For instance, the Holzapfel-Gasser-Ogden model for arteries\cite{HoGO00} is now implemented in many Finite Element Analysis software packages.
It requires the experimental determination of only 3 elastic constants per layer of arterial wall.


\section*{Acknowledgments}

This work is supported by a Senior Marie Curie Fellowship awarded by the Seventh Framework Programme of the European Commission to the first author and by an E.T.S. Walton Award, given to the third author by Science Foundation Ireland.  This material is based upon works supported by the Science Foundation Ireland under Grant No. SFI 08/W.1/B2580.


\end{document}